# A Comparative Analysis Towards Melanoma Classification Using Transfer Learning by Analyzing Dermoscopic Images

Md. Fahim Uddin[1], Nafisa Tafshir[2], Mohammad Monirujjaman Khan[1,*]

[1]Department of Electrical and Computer Engineering,
North South University, Bashundhara, Dhaka-1229, Bangladesh

Mohammad Monirujjaman Khan. Email: monirujjaman.khan@northsouth.edu

**Abstract:** Melanoma is a sort of skin cancer that starts in the cells known as melanocytes. It is more dangerous than other types of skin cancer because it can spread to other organs. Melanoma can be fatal if it spreads to other parts of the body. Early detection is the key to cure, but it requires the skills of skilled doctors to diagnose it. This paper presents a system that combines deep learning techniques with established transfer learning methods to enable skin lesions classification and diagnosis of melanoma skin lesions. Using Convolutional Neural Networks, it presents a method for categorizing melanoma images into benign and malignant images in this research (CNNs). Researchers used 'Deep Learning' techniques to train an expansive number of photos & essentially to get the expected result deep neural networks to need to be trained with a huge number of parameters as dermoscopic images are sensitive & very hard to classify. This paper, has been emphasized building models with less complexity and comparatively better accuracy with limited datasets & partially fewer deep networks so that the system can predict Melanoma at ease from input dermoscopic images as correctly as possible within devices with less computational power. The dataset has been obtained from ISIC Archive. Multiple pre-trained models ResNet101, DenseNet, EfficientNet, InceptionV3 have been implemented using transfer learning techniques to complete the comparative analysis & every model achieved good accuracy. Before training the models, the data has been augmented by multiple parameters to improve the accuracy. Moreover, the results are better than the previous state-of-the-art approaches & adequate to predict melanoma. Among these architectures, DenseNet performed better than the others which gives a validation accuracy of 96.64%, validation loss of 9.43% & test set accuracy of 99.63%.

## 1 Introduction

One of the foremost perilous sorts of cancer is skin cancer Different sorts of skin cancer can be found in different parts of the body. Melanoma is the most unpredictable skin cancer.[1] According to the most recent WHO data, skin cancer fatalities in Bangladesh reached 301 in 2018, accounting for 0.04 percent of all deaths. Bangladesh ranks #183 in the world with an age-adjusted Death. [2]

According to WHO figures, there are 324,635 new cases worldwide, with 57,043 of those being death cases. It was discovered that 18 persons out of 100 who were diagnosed with Melanoma died, regardless of whether the cancer was considered curable [3]. It is, without a question, disturbing and a source of concern for dermatologists all over the world.

One in three white men and one in 40 white women will be affected by melanoma in their lifetime. (a woman on 156 against one out of 230 men). Overall, a white man on 27 and a white woman on 40 will develop melanoma in their life [4]

This cancer is the most common cancer in the US, with over 5 million cases diagnosed every year [5]. Each year, approximately 5.4 million new skin cancer cases are recorded in the US alone There are over 13,000 new instances of melanoma yearly, leading to over 1,200 deaths in Australia [6]. Each year, melanoma causes over 20,000 deaths in Europe [7]. Global statistics are just as good. According to recent reports, the incidence of melanoma has increased significantly over the last three decades, with an estimated more than 96,000 new cases diagnosed in the US in 2019. society. The mortality rate of this disease increases during the next decade is expected. When diagnosed late, the survival rate is less than 14%. However, if skin cancer is detected early, the survival rate is close to 97%. [8] This requires early location of skin cancer. This paper improves precision and solves the problem of early diagnosis.

Skin cancer is extremely frequent in Europe, Australia, and US, and it is almost always curable if caught and treated early. Skin color, sun exposure, climate, old age, genetic and familial history are all key risk factors. Melanoma can be identified by recognizing a unused spot on the skin or one that's changing in measure, shape, or color. Early recognizable proof of skin cancer can spare a person's life.. [9]

The use of deep learning and machine learning in computer-aided diagnosis systems has become a prerequisite for the early detection and diagnosis of many fatal diseases in today's world. Our technology also functions as a computer-aided diagnosis system, [10] determining whether or not an image of damaged skin contains melanoma cancer. Deep layered convolutional network algorithms were used to accomplish this. CNN's deep layers train the data set and extract features with greater ease and accuracy. Picture processing was also done to reduce noise from image data and make it easier and more suited for training in the model. According to **Fig 1** since 1974 melanoma incidence rate is increasing gradually & this rate is significantly higher in recent years. [11]

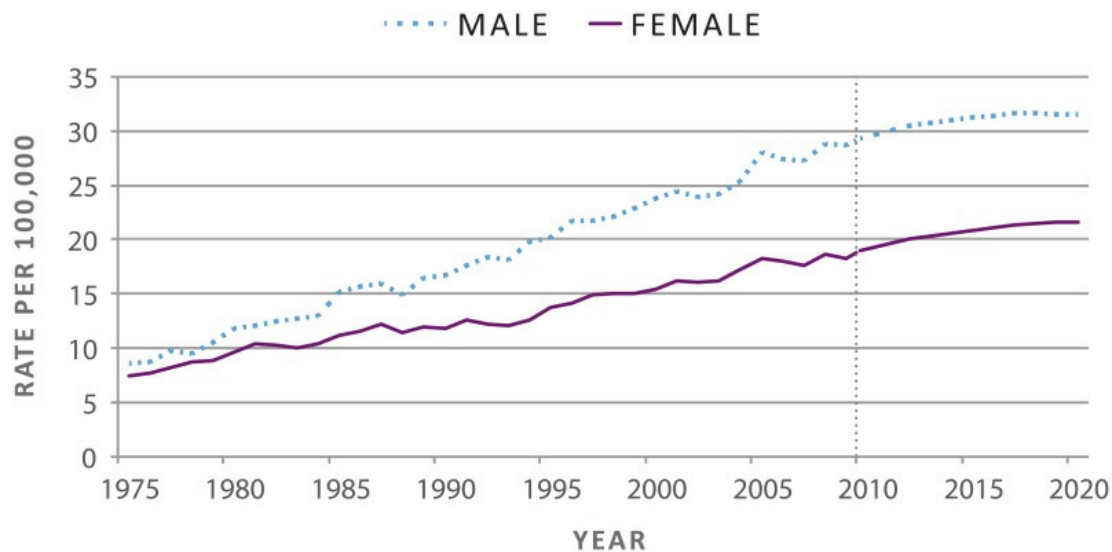

**Figure 1:** Melanoma Incidence Rate Based on Age Sex from 1975-2020 [11]

In recent years, usage of deep CNN in medical image classification is increasing gradually & these networks have shown promising results. Earlier in the research of melanoma detection, deep learning algorithms were vastly used.

**Arief et al.** used the ResNet model for classifying melanoma cancer and normal skin images. The architecture was learned using augmented and under sampled data trains. The F1 Score was used to obtain the validation outcome for each model. Their best- performing architecture ResNet50 gave validation accuracy of 83% [13]. **Alex et al.** wrote a highly optimized GPU usage of 2D convolution and all the other operations inherent in preparing convolutional neural systems. They utilized non-saturating neurons and an awfully proficient GPU implementation of the convolution prepare to create training go speedier. They used a recently invented regularization method termed "dropout" to decrease overfitting within the completely associated layers, which proved to be quite effective. Their architecture AlexNet gave validation accuracy of 73.5% [14]. **Karen et al.** came up with altogether more precise ConvNet structures, which not as it were accomplish the state-of-the-art accuracy on ILSVRC classification and localization errands but are too appropriate to other picture acknowledgment datasets, where they accomplish amazing performance indeed when utilized as a portion of generally straightforward pipelines (e.g. deep features classified by a linear SVM without fine-tuning). Their architecture VGG16 gave validation accuracy of 72.6% [15]. **Ahmet et al.** used ResNet-101 and Inception-v3 neural network architectures for early diagnosis of skin cancer by classifying their dataset images as benign or malignant. In the ResNet-101 architecture, they achieve an accuracy rate of 84.09 %, and in the Inception-v3 architecture, they get an accuracy rate of 87.42 %. [16]. **Md. Fazle et al.** used ResNet50, InceptionV3, Xception, and VGG16 for skin cancer classification and three state-of- the-art techniques SegNet, BCDU-Net, and U-Net [19] for lesion segmentation. Their architecture Xception gave validation accuracy of 95.2% [17]. **Codella et al.** proposed a hybrid approach, integrating convolutional neural network (CNN), sparse coding and support vector machines (SVMs) to detect melanoma [20]. We tabulate all the accuracy of the mentioned studies on Melanoma detection in Table 6.

In the field of skin cancer classification utilizing deep learning and computer vision techniques, there has been various types of research published. These studies employ a variety of techniques, including classification only, segmentation and detection, image processing using various types of filters, and so on.

This paper developed an automated diagnosis system for melanoma classification utilizing dermoscopy images by using multiple pre-trained models & transfer learning methods. For melanoma identification, this paper used some parameters related to form, size, and color attributes to differentiate the benign & malignant skin lesions. The model is trained using the ResNet101, DenseNet, EfficientNet, InceptionV3. The methods and materials utilized to design and construct the system are highlighted in the subsequent areas of this paper, and the results are analyzed in Section 2. In Table 3, we even illustrated the accuracy and usefulness of several earlier research on Melanoma Detection.

## 2 Methods and Materials

The dataset was taken from the open-source ISIC Archive to train & test the model. The dataset contained dermoscopic images of both benign & malignant skin lesions. Data augmentation is also used to get the best result of this research. In this research, different types of Convolutional Neural Network (CNN) like ResNet, DenseNet, InceptionV3, EfficientNet have been used & transfer learning techniques are used for categorization with the use of pre-trained models. A pre-trained model has been trained on a big benchmark dataset to handle a problem that is comparable to the one we are working on. As a result, the model uses patterns learned while addressing a separate problem. In this way, the model builds on existing knowledge rather than starting from scratch.

### 2.1 Materials and Tools

For data analysis, Python was the best programming language. Because of Python's broad library access, deep learning-based challenges are especially compelling with Python programming. Expansive datasets and model training were handled online using Google Colab. All data, code, and work were saved to GitHub so that they could be retrieved from any GPU. GitHub is appropriate for teamwork because it offers a tracking mechanism for collaborative and code management.

*2.2 Dataset Description*

The International Symposium on Biomedical Imaging (ISBI) raises problems in several biomedical sectors every year. Skin cancer detection is one of the difficulties. The dataset we used for skin cancer identification was from a competition held in 2016 and was available on the ISIC [16] website. We took 10000 dermoscopic images where Benign is 5000 and Malignant is 5000. We utilized 3500 photos of each to train our model of each benign and malignant class and 1500 images for our testing. Dataset Link: [ISIC Archive](#) [16]

This page contains photographs of various sizes, which we must resize. We reduce the size of our photograph to (224, 224, 3). As a result, the contour of our input image is (224, 224,3). Here's a sample of the photos we took for this project:

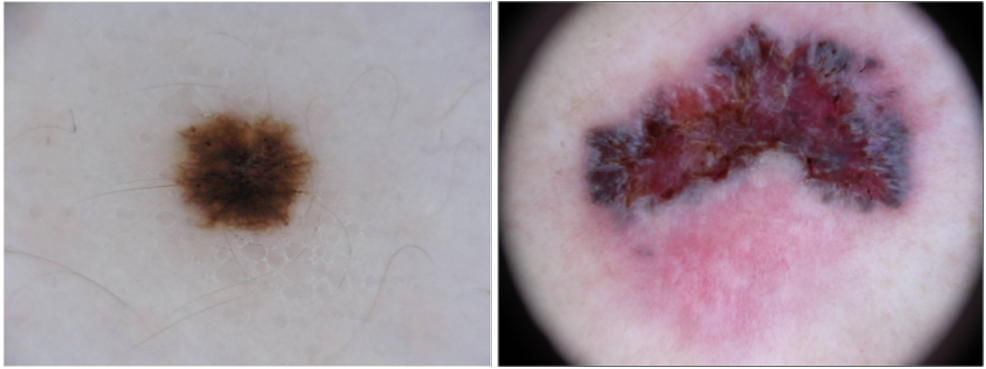

**Figure 2:** Dermoscopic Image of Benign & Malignant Skin Lesion

We took 3500 for training & 1500 for validation for both benign & malignant. So, over all the percentage of the training dataset & validation dataset is 70% & 30%.

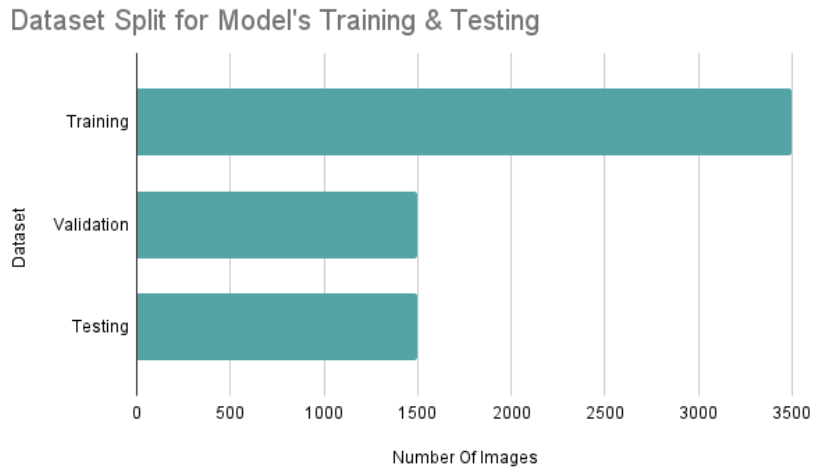

**Figure 3:** Dataset Split for Model's Training & Testing

The size images of melanoma we collected from ISIC are very big. We have resized our Dataset image size to make the training faster. We resized our dataset images 224 px * 224 px. We took 3500 for training & 1500 for validation for both benign & malignant. So, over all the percentage of the training dataset & validation dataset is 70% & 30%.

*2.3 Data Augmentation*

Deep learning-based algorithms are being studied in medical image classification. Their ability to outperform traditional machine learning methods has become more prevalent Data augmentation refers to the process of increasing the amount of data that is needed to work properly with a limited dataset. Our research is based on a limited number of medical datasets which contain images of skin lesions. So, in **Tab 1** parameters have been used for the data augmentation to get the best result of the model. So, in **Tab 1** parameters have been used for the data augmentation to get the best result of the model.

**Table 1:** Data Augmentation Parameters

| Parameters | Value |
| --- | --- |
| Zoom Range | 2 |
| Rotation Range | 90 |
| Horizontal Flip | True |
| Vertical Flip | True |

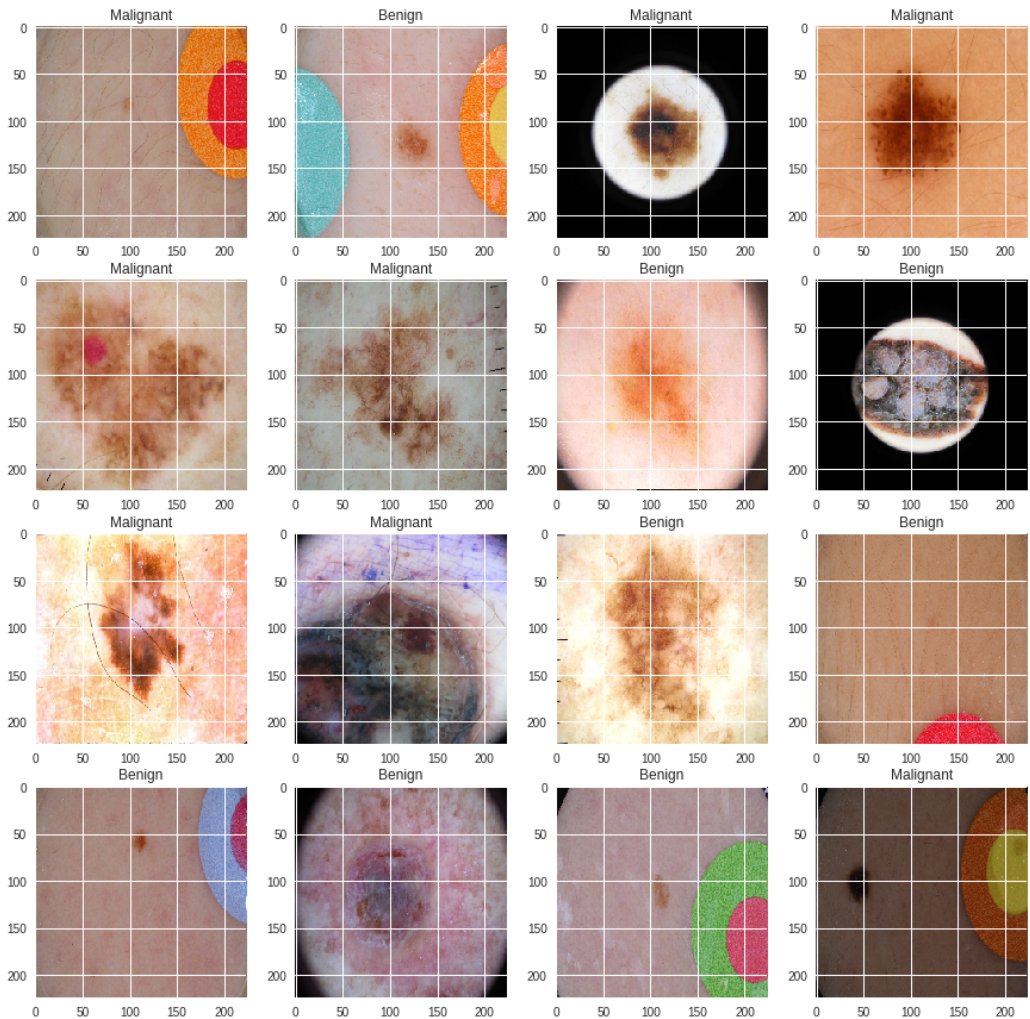

**Figure 4:** Random images of the dataset after augmentation

## *2.4 Block Diagram*

In the below **Fig. 5**, As we said before that after preprocessing data, we part the data into two portions. 70% is used to train the model & 30% is used to test the model. Before training the model, we have labeled our dataset. We labeled class 0 for our benign images & class 1 for our malignant images. So, the model can differentiate benign and malignant. After training the model, the new data is predicted in the trained model. Afterward, the model gives the predicted result of whether the tumor is benign or malignant.

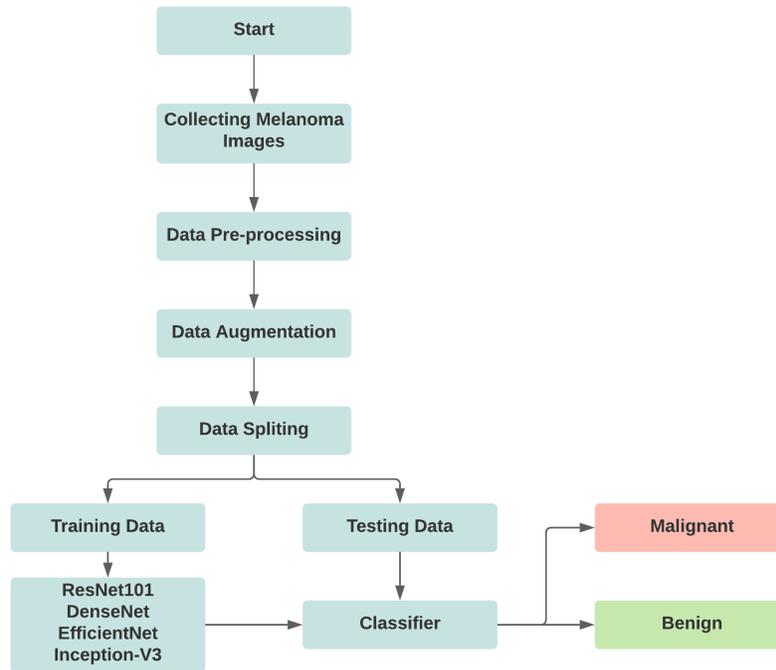

**Figure 5:** Block diagram of the proposed method

*2.5 System Architecture*

    The system architecture is the overview of the whole system. In this architecture, During the training of our model, we have completed visualization of the dataset as the images of the dataset would be larger. Then we part the data into train & test. The percentage of train & test is as follows: Train - 70% & Test - 30%. After all this preprocessing, the features enter into the network**. Fig. 6** shows a diagram view of the architecture.

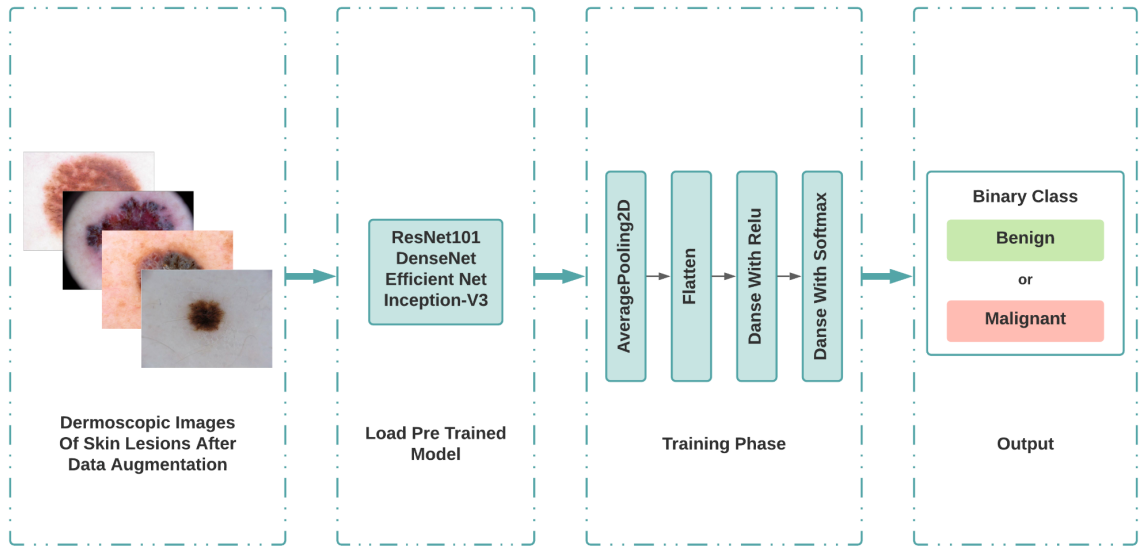

**Figure 6:** System architecture of the pre-trained model

*2.5.1 Convolutional layer*

    CNN are a sort of neural network which is extremely effective at picture recognition and classification. CNN are utilized in real-time applications such as robots and self-driving automobiles because they can recognize faces, pedestrians, traffic signs, and other things better than humans. CNN are supervised learning methods that are trained using data that has been labeled with the appropriate classes. CNN learns the relationship between input objects and class labels, including two parts: a hidden layer for extracting features, and a completely associated layer for actual classification tasks at the conclusion of processing. CNN's hidden layer has a specific architecture, consisting of a convolutional layer, a grouping layer, and a trigger function to turn neurons on or off. In a typical neural network, each layer is made up of a group of neurons, and one neuron in one layer is associated to each neuron within the previous layer. The hidden layer in CNN is different from the previous layer. Instead of having multiple connections between neurons, the new neurons only have one connection to each other. Due to fewer parameters and less complexity of the model, this leads to a simpler training process.

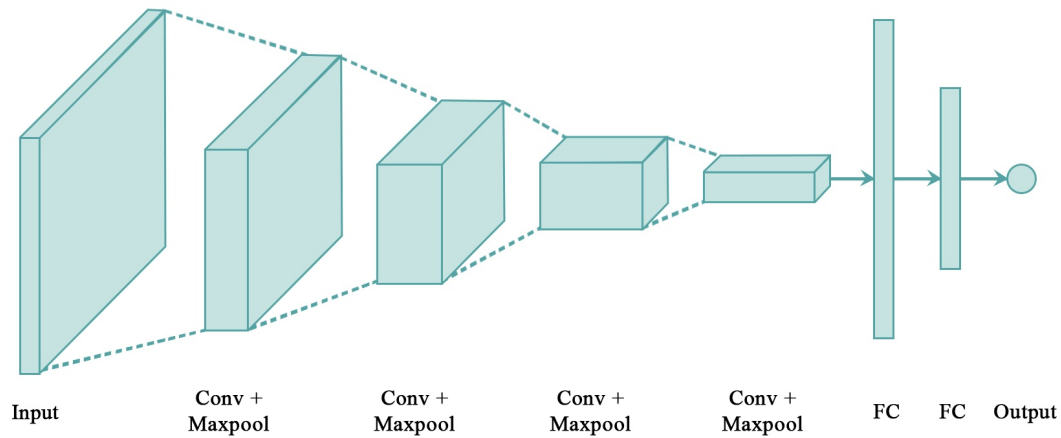

**Figure 7:** Layers of CNN

*2.5.2 Pretrained Model*

    Not having enough medical data is a major challenge for researchers in the field of medical research. This issue can be solved through deep learning methods. Transfer learning is a method that eliminates the need for large data sets. It saves money and time by transferring data from a pre-trained model to a new model that needs training with smaller datasets Four CNN-based models were used for analyzing Dermoscopic images. The models were ResNet101, InceptionV3, EfficientNet., and DenseNet169

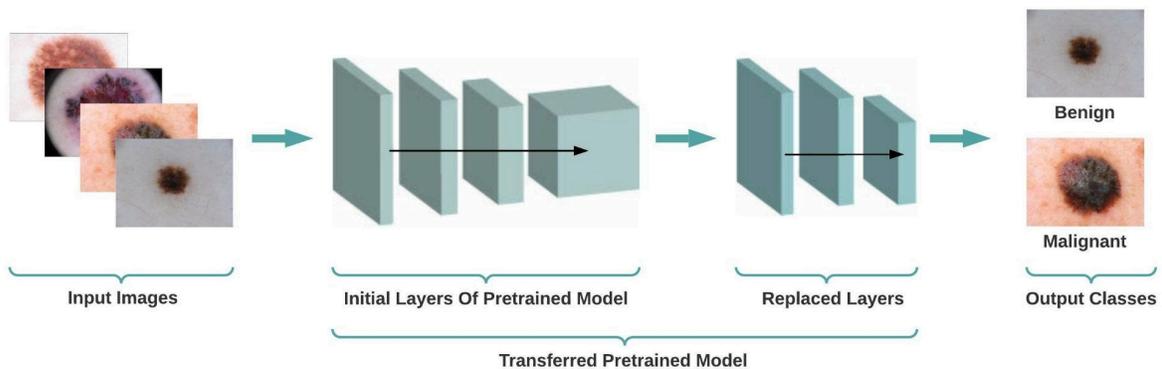

**Figure 8:** Diagram of the pre-trained model

*2.5.3 ResNet*

    ResNet is short for Residual Network, which defines "residual learning" as a key term introduced by the network. ResNet-101 is a 101-layer Residual Network that is a modification of ResNet-50. ResNet's main innovation is skipping connections. As you may know, constant deep networks often encounter the issue of vanishing gradients, that is, as the model propagates backward, the slope becomes smaller and smaller. Small gradients can make learning difficult to manage. The jump connection in the diagram

below is called "identity". It allows the network to be proficient in the identity function, allowing it to pass input through squares without passing through other weight layers. Therefore, you can stack additional layers and build a deeper network by compensating for the disappearing gradient, so that our network can overcome the complexity of training.

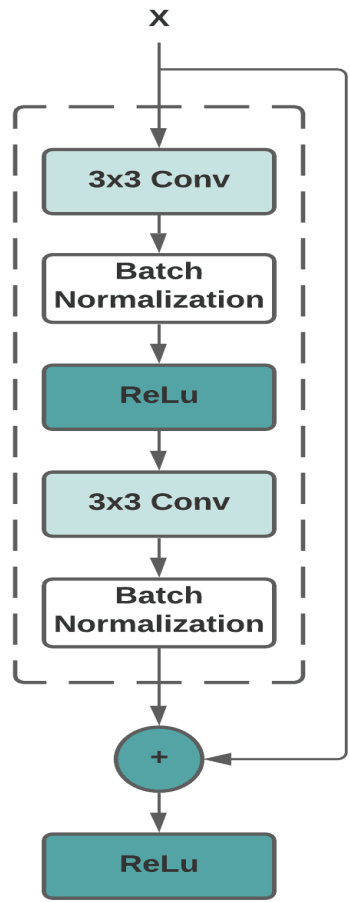

**Figure 9:** Single residual block [11]

*2.5.4 DenseNet*

DenseNet is a new CNN architecture that uses fewer parameters to achieve State-Of-The-Art (SOTA) performance on classification datasets (CIFAR, SVHN, ImageNet). It can be more profound than traditional networks while still being straightforward to optimize. The DenseNet system is made up of Dense blocks. The layers of those blocks are extensively interconnected: All preceding layers' output feature maps are fed into each layer. By connecting every layer (of the same dense block) to its succeeding levels, the DenseNet design maximizes the residual mechanism. Because the acquired features are all shared through common knowledge, the model's compactness makes them non-redundant. It's also a lot easier to train a deep network with dense connections thanks to implicit deep supervision where the gradient is used.

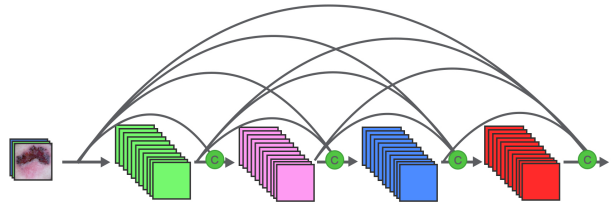

**Figure 10:** Dense Connectivity

*2.5.5 EfficientNet*

Engineering and scale are at the heart of EfficientNet. It demonstrates that by properly designing your architecture, you can achieve excellent results while working within sensible constraints. Images can be organized into 1,000 objects, such as keyboards, various animals, mice, and pens, via the network. As a result, the network gained the ability to represent various images in a multifunctional manner.

*2.5.6 Inception-V3*

A 42-layer deep neural network makes up the Inception-v3 model. Convolutions, max-pooling layers, normal pooling, dropouts, and completely connected layers are among the symmetric and asymmetric building components within the Inception-v3 model. It may be a commonly utilized picture acknowledgment demonstrate that has been appeared to achieve an exactness rate of more noteworthy than 78.1% on the ImageNet dataset. The model represents the result of several ideas explored over time by a number of researchers.

*2.6 Evaluation Metrics*

The execution of the proposed demonstrate is based on different metrics: accuracy, recall, sensitivity, specificity, and precision. Metrics are evaluated by various parameters in the confusion matrix, such as true positive (TP), true negative (TN), false positive (FP), and false-negative (FN).

*2.6.1 Confusion Matrix*

It is a matrix that describes the performance of an algorithm. The confusion matrix contains information about the actual class and predictions of the model [18].

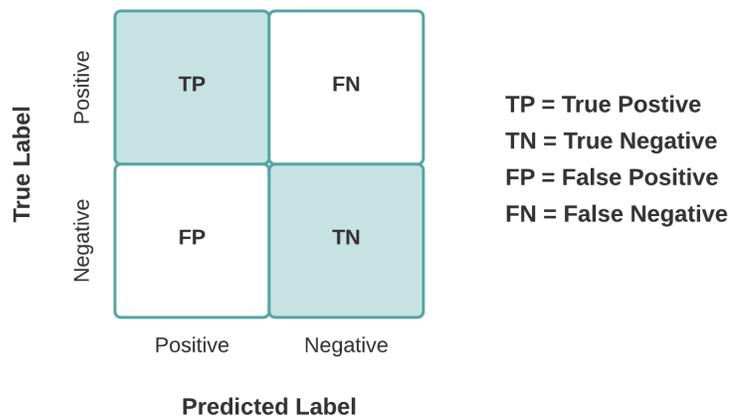

**Figure 11:** Confusion matrix

*2.6.2 Recall, Precision, F1 Score*

The accuracy of the classifier is quantified with this metric. The number of correctly classified data is divided by the total number of the data to calculate accuracy.

$$\text{Accuracy} = \frac{TP+TN}{TP+TN+FP+FN} \tag{1}$$

Precision shows how much of the data predicted as positive are predicted correctly. To put it another way, greater precision implies fewer false positives.

$$Precision = \frac{TP}{TP+FP} \tag{2}$$

The recall is the metric of determining the completeness of the classifier. Higher recall indicates lower false negatives, while lower recall indicates higher false negatives. Precision often decreases with an improvement in recall.

$$Recall = \frac{TP}{TP+FN} \tag{3}$$

To obtain the F1-score, the product of recall and precision is divided by the sum of recall and precision.

$$F1 - \text{Score} = 2 \times \frac{Recall \times Precision}{Recall + Precision} \tag{4}$$

**3 Result & Analysis**

After data augmentation, we trained each model with a trained generator & validation generator for 20 epochs and we achieved good model accuracy for every model. The training factors have been tabulated in Tab 2 and have been followed for each & every model.

**Table 2:** List of Materials and Tools

| Training Factor | Values |
|---|---|
| Platform | Google COLAB |
| GPU | Colab GPU (NVIDIA Tesla K80) |
| Optimizer | Adam |
| Loss Function | Categorical Cross-Entropy |
| Learning Rate | 0.0001 |
| Epoch | 20 |
| Batch Size | 64 |

*3.1 Model Evaluation*

The training accuracy increased significantly after each epoch, as can be seen from the plot of accuracy history. ResNet101 achieved validation accuracy of 95.71% & validation loss of 12.73%. DensNet169 achieved validation accuracy of 96.64% & validation loss of 9.43%. EfficientNet achieved validation accuracy of 96.07% & validation loss of 9.78%. InceptionV3 achieved validation accuracy of 96.50% & validation loss of 9.33%. So, among all these models DenseNet achieved the highest validation accuracy & lowest validation loss. We have tabulated the validation accuracy & validation loss in **Tab 3**

| Model | Validation Accuracy | Validation Loss |
|---|---|---|
| ResNet101 | 0.9571 | 0.1273 |
| DenseNet169 | 0.9664 | 0.0943 |
| EfficientNet | 0.9607 | 0.0978 |
| InceptionV3 | 0.9650 | 0.0933 |

We have tabulated the test set accuracy in **Tab 4**. Among all the models, ResNet101 & DenseNet169 achieved the highest accuracy by obtaining an accuracy of 99.63%. EfficientNet & InceptionV3 obtained were close to them by obtaining an accuracy of 99.16%. But DenseNet169 performed better than ResNet101 in the validation set.

**Table 4:** The Test Set Accuracy of The Models

| Models | Test Set Accuracy |
|---|---|
| ResNet101 | 0.9963 |
| DenseNet | 0.9963 |
| EfficientNet | 0.9916 |
| Inception V3 | 0.9916 |

In **tab 5** among all the models ResNet101 & DenseNet169 achieved the highest F1 Score by obtaining 100%. EfficientNet & InceptionV3 achieved a F1 score of 99%. So, in terms of F1 score ResNet101 & DenseNet169 gave better results.

**Table 5:** The Recall, Precision, And Accuracy of The Models

| Model | Class | Precision | Recall | F1 Score |
|---|---|---|---|---|
| ResNet101 | Benign | 1.00 | 1.00 | 1.00 |
| | Malignant | 1.00 | 1.00 | 1.00 |
| DenseNet169 | Benign | 0.99 | 1.00 | 1.00 |
| | Malignant | 1.00 | 0.99 | 1.00 |
| EfficientNet | Benign | 0.98 | 1.00 | 0.99 |
| | Malignant | 1.00 | 0.98 | 0.99 |
| InceptionV3 | Benign | 0.98 | 1.00 | 0.99 |
| | Malignant | 1.00 | 0.98 | 0.99 |

By analyzing Validation accuracy, validation loss & F1 score it can be stated that DensNet169 performed better than ResNet101, EfficientNet & InceptionV3. ResNet101 gave almost close but the validation loss is 25% higher than DensNet169. So, DenseNet169 is better than other models in terms of all evaluation matrices.

Anyone can see that train accuracy has increased rapidly after each epoch by looking at the accuracy graph. The accuracy was 85.75 % in the first epoch, then increased with each epoch. In comparison, the model's validation accuracy was 94.36 %, and it continued to increase until the last epoch. It can be seen on the plot of model accuracy that an increasing line has been established for train accuracy and test accuracy; it has plotted a line which is around the region of 94%-99% accuracy all the time during the epoch. Figs. show the model accuracy and model loss.

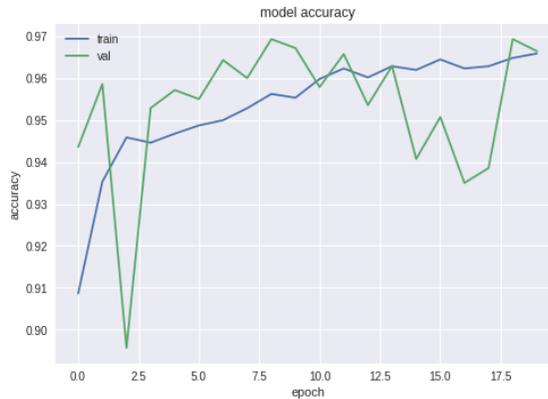

**Figure 12:** Model accuracy of DenseNet169

People can understand from the model loss graph that both the training and test loss line have gradually decreased. The train loss was 32.97 % after the first epoch and 11.01 % after 10 epochs. Validation loss was 21.29 % after 10 epochs, while it was 11.59 % after 10 epochs. A plot of model loss is shown in Fig.

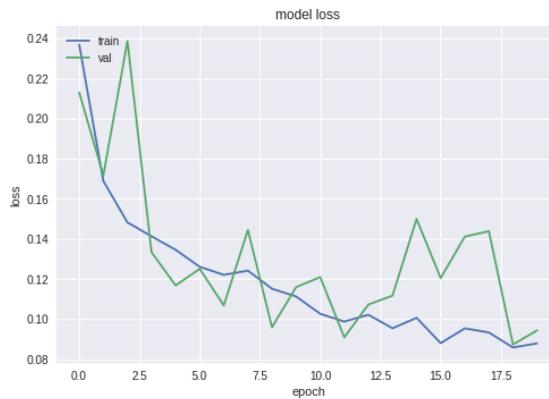

**Figure 13:** Model loss of DenseNet169

The accuracy of all algorithms is kind of the same, but DenseNet has given better results from the start of training. Of all those architectures, DenseNet169, and ResNet101 have given better results compared to others in terms of accuracy and training loss. After the testing of the model, it also detected Benign and Malignant correctly.

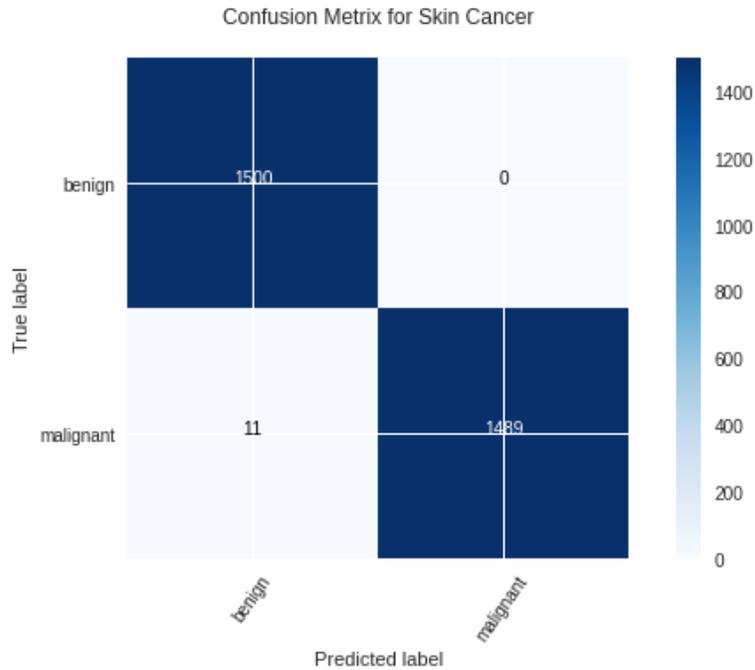

**Figure 14:** DenseNet169 confusion matrix

*3.2 Model Test*

This study also included real testing by giving dermoscopic images of skin as input to the model. When the model is ready. After that, a new notebook file is created as an ipynb extension for the test. In that test file, four models were included and then random dermoscopic images from the test set were provided as input. In Fig. Below, the result of the test has shown the prediction whether it is Covid or not.

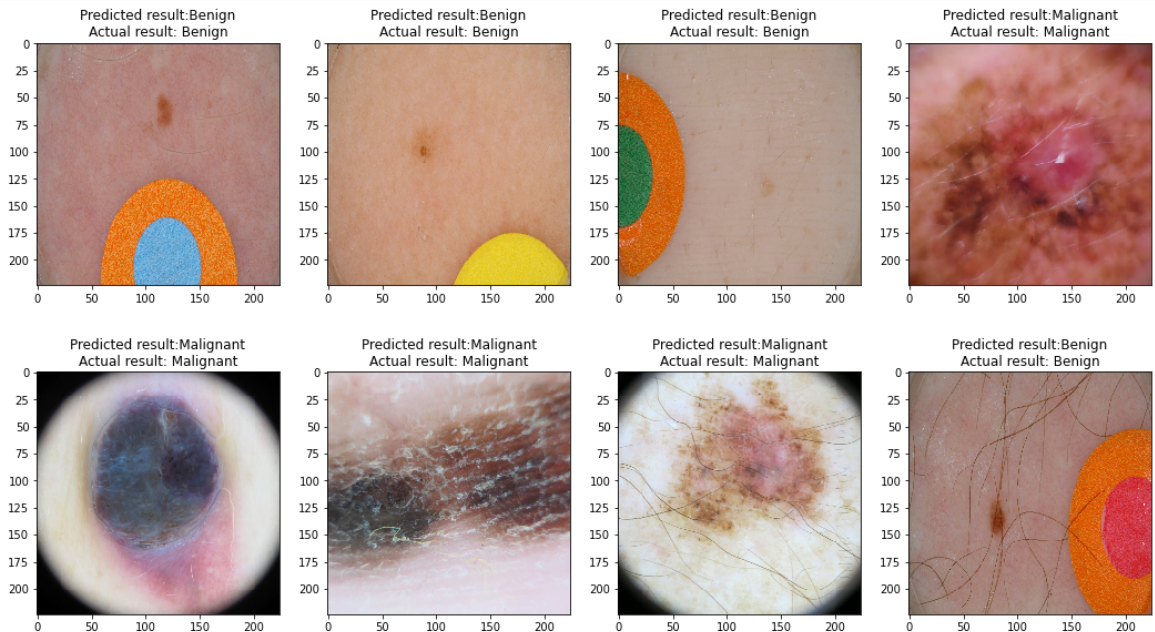

**Figure 15:** Model Testing

After research and analysis, the results show that in almost all models used in this study, the migration learning model of DenseNet169 is superior to and exceeds the accuracy of all experiments. The scores are shown in the **Tab 5** According to observations, they have correctly classified the dermoscopic images. From the overall results and analysis, this paper found that all algorithms work well in the model and give an accuracy of more than 98%. Using CNN, which is more accurate than ordinary deep learning algorithms, is the author's main contribution, which creates a basic platform for other researchers to continue their research in these types of fields. Therefore, this research is very effective for maintaining sustainable development goals.

*3.3 Model Comparison*

Pre-trained models of this paper have been compared with some previous models. Compared with the above referenced papers, ResNet101, DenseNet169, EfficientNet and InceptionV3 have given us better results, accuracy of networks and efficiency in this paper. In ResNet101 & InceptionV3 pre-trained models, accuracy has increased to a significant level.

**Table 6:** Test Set Accuracy of Other Papers for Comparative Analysis

| Reference | Architecture | Validation Acc |
|---|---|---|
| In study [13] | ResNet 50 | 0.785 |
| | ResNet 40 | 0.835 |
| | ResNet 25 | 0.807 |
| | ResNet 10 | 0.822 |
| | ResNet 7 | 0.824 |
| In study [14] | AlexNet | 0.735 |
| In study [15] | VGG16 | 0.726 |
| In study [16] | ResNet 101 | 0.890 |
| | InceptionV3 | 0.900 |
| In study [17] | ResNet 50 | 0.933 |
| | Xception | 0.952 |
| | VGG16 | 0.931 |
| | Inception V3 | 0.941 |

Prediction based error analysis can be done using a confusion matrix where it can be visualized by the percentage of true positive, true negative, false positive, and false negative. Data size and data nature are also important for error analysis. Splitting the data accordingly for making trains and tests is also considerable for error analysis as the train and test sets may affect the result on a large scale. Features play an important role in error analysis. Feature engineering and regularization is also done for reducing the errors.

**4 Conclusion**

Melanoma is a malignancy that is difficult to detect conventionally. Apart from being a person with melanoma cancer who is painless, the form of melanoma cancer is also comparable to regular moles. The damage to DNA in melanoma cancer is caused by overexposure to ultraviolet rays (UV), and the damaged cells are melanocytes, which create melanin (pigmentation of the skin). Finally, this research will look into the potential of deep convolutional neural networks to differentiate between benign and malignant skin cancer. We show that with the utilize of exceptionally profound convolutional neural systems utilizing exchange learning and fine-tuning them on dermoscopy pictures, way better demonstrative precision can be accomplished compared to master doctors and clinicians if a cost-effective system can be built, it can easily be reachable to general people and melanoma screening can be easier. And yes, saving lives is our ultimate goal. In this study, the objective is to get the best model for classifying melanoma cancer and normal skin images. This paper presents a comparative study for melanoma detection with CNN. To build a cost-effective model, we are presenting the knowledge distillation method. A way of exchanging information from one arrange to another, with experimental results over several images of Melanoma datasets, the main goal is to save lives from diseases like Melanoma with the help of new technology. This technology will inspire future generations to cope with this kind of unwanted situation. In the future, Image data from a massive number of melanoma patients can be included in the dataset, and training those can be a great way to see how outcomes change. Other networks of CNN models can be trained to see the accuracy values and compare them in the context of accuracy, precision, recall, and F1 score.

**Data Availability Statement**

The data used to support the findings of this study are freely available at https://challenge.isic-archive.com/

**Conflicts of Interest**

The authors would like to confirm there are no conflicts of interest regarding the study.

**Acknowledgement**

The author would like to thank DR.Mohammad Monirujjaman Khan, Department of Electrical and Computer Engineering, North South University, Dhaka, Bangladesh, for supporting this research. We also thank the reviewers for their comments and suggestions in advance.